\documentclass[sigplan,nonacm]{acmart}
\usepackage{xspace}
\AtBeginDocument{%
  }

\setcopyright{acmlicensed}
\copyrightyear{2018}
\acmYear{2018}
\acmDOI{XXXXXXX.XXXXXXX}
\acmConference[Conference acronym 'XX]{Make sure to enter the correct
  conference title from your rights confirmation email}{June 03--05,
  2018}{Woodstock, NY}
\acmISBN{978-1-4503-XXXX-X/2018/06}

\begin{document}

\title{AscendCraft: Automatic Ascend NPU Kernel Generation via DSL-Guided Transcompilation}

\author{Zhongzhen Wen}
\affiliation{%
  \institution{State Key Lab for Novel Software Technology, Nanjing University}
  \city{Nanjing}
  \country{China}
}
\email{wenzhongzhen@smail.nju.edu.cn}

\author{Shudi Shao}
\affiliation{%
  \institution{Software Engineering Application Technology Laboratory, Huawei}
  \city{Shanghai}
  \country{China}
}
\email{shaoshudi@huawei.com}

\author{Zhong Li}
\affiliation{%
  \institution{State Key Lab for Novel Software Technology, Nanjing University}
  \city{Nanjing}
  \country{China}
}
\email{lizhong@nju.edu.cn}

\author{Yu Ge}
\affiliation{%
  \institution{State Key Lab for Novel Software Technology, Nanjing University}
  \city{Nanjing}
  \country{China}
}
\email{yuge@smail.nju.edu.cn}

\author{Tongtong Xu}
\affiliation{%
  \institution{Software Engineering Application Technology Laboratory, Huawei}
  \city{Hangzhou}
  \country{China}
}
\email{xutongtong9@huawei.com}

\author{Yuanyi Lin}
\affiliation{%
  \institution{Software Engineering Application Technology Laboratory, Huawei}
  \city{Hangzhou}
  \country{China}
}
\email{linyuanyi2@huawei.com}

\author{Tian Zhang}
\affiliation{%
  \institution{State Key Lab for Novel Software Technology, Nanjing University}
  \city{Nanjing}
  \country{China}
}
\email{ztluck@nju.edu.cn}

\renewcommand{\shortauthors}{Wen et al.}
\newcommand{\ourtool}{\textsc{AscendCraft}\xspace}

\begin{abstract}
The performance of deep learning models critically depends on efficient kernel implementations, yet developing high-performance kernels for specialized accelerators remains time-consuming and expertise-intensive. While recent work demonstrates that large language models (LLMs) can generate correct and performant GPU kernels, kernel generation for neural processing units (NPUs) remains largely underexplored due to domain-specific programming models, limited public examples, and sparse documentation. Consequently, directly generating AscendC kernels with LLMs yields extremely low correctness, highlighting a substantial gap between GPU and NPU kernel generation.

We present \ourtool, a DSL-guided approach for automatic AscendC kernel generation. \ourtool introduces a lightweight DSL that abstracts non-essential complexity while explicitly modeling Ascend-specific execution semantics. Kernels are first generated in the DSL using category-specific expert examples and then transcompiled into AscendC through structured, constraint-driven LLM lowering passes. Evaluated on MultiKernelBench across seven operator categories, \ourtool achieves 98.1\% compilation success and 90.4\% functional correctness. Moreover, 46.2\% of generated kernels match or exceed PyTorch eager execution performance, demonstrating that DSL-guided transcompilation can enable LLMs to generate both correct and competitive NPU kernels. Beyond benchmarks, \ourtool further demonstrates its generality by successfully generating two correct kernels for newly proposed mHC architecture, achieving performance that substantially surpasses PyTorch eager execution.
\end{abstract}





\maketitle

\section{Introduction}~\label{sec:intro}

The performance of deep learning (DL) models critically depends on the efficiency of underlying DL kernels.
Developing high-performance kernels is a time-consuming and expertise-intensive process, requiring deep understanding of hardware architectures, memory hierarchies, and low-level optimization techniques.
This challenge is particularly evident on specialized accelerators, where achieving high performance often requires careful orchestration of data movement and on-chip resources.

Recent advances in large language models (LLMs) for code generation have attracted significant attention from the research community, including efforts to automatically generate DL kernels.
Most existing work focuses on GPU kernels, benefiting from abundant public corpora, extensive documentation, and mature programming ecosystems.
As a result, prior studies~\cite{baronio2025kevinmultiturnrlgenerating, wei2025astra, ouyang2025kernelbenchllmswriteefficient} have demonstrated promising results in generating correct and performant GPU kernels using LLMs.

In contrast, kernel generation for neural processing units (NPUs) remains largely unexplored.
NPU programming models are often domain-specific, with limited public examples and sparse documentation.
As a result, LLMs lack sufficient exposure to NPU-specific syntax, execution models, and optimization principles.
Recent evidence from MultiKernelBench~\cite{wen2025multikernelbenchmultiplatformbenchmarkkernel} shows that state-of-the-art LLMs achieve a correctness rate below 5\% when directly generating AscendC kernels, highlighting the substantial gap between GPU and NPU kernel generation.

To bridge this gap, we propose \ourtool, a DSL-guided approach for automatic AscendC kernel generation.
At the core of \ourtool is a lightweight domain-specific language (DSL) designed around three key principles.
\textbf{(1) Concise and structured syntax:}
the DSL adopts a compact and regular programming structure that reduces syntactic verbosity and enforces clear control flow.
This design allows the model to focus on core algorithmic decisions—such as tiling strategies and dataflow organization—rather than low-level syntactic details, while keeping programs short and stable for generation.
\textbf{(2) Appropriate abstraction:}
the DSL deliberately hides platform-specific complexities that impose significant programming burden but are not essential for expressing core kernel logic.
For example, unaligned memory access handling in AscendC requires verbose \texttt{DataCopyPad} configurations with numerous arguments; such details are abstracted away in the DSL to simplify generation without sacrificing correctness.
\textbf{(3) Targeted extensions:}
the DSL explicitly models Ascend-specific execution semantics, including on-chip memory allocation (e.g., Unified Buffer usage), staged execution (\texttt{CopyIn, Compute, CopyOut}), and inter-core execution structure.
These targeted extensions ensure that the DSL remains expressive enough to capture high-performance NPU kernel designs.

Our approach consists of two stages.
In the first stage, the LLM generates a high-level DSL program that describes the kernel’s core computation, tiling strategy, and on-chip dataflow.
The DSL examples provided to the LLM are written by human experts and are \textbf{category-specific}, with each category corresponding to a class of operators that share similar computation patterns and optimization objectives.
These examples encode common optimization strategies and execution patterns for Ascend kernels within the same operator category.
By operating on this \textbf{concise and structured DSL}, the LLM can effectively learn category-level core design principles and generalize them to unseen operator configurations within the same category.

In the second stage, we transcompile the DSL into AscendC code through a sequence of \textbf{structured LLM-based lowering passes}, each responsible for translating a specific aspect of the DSL into valid and efficient AscendC constructs.
\textbf{Decomposing} the translation process into well-defined subtasks reduces the complexity of each generation step, thereby improving both correctness and robustness.
Moreover, we \textbf{explicitly constrain the structure} of the generated AscendC program during transcompilation.
In particular, each \texttt{copyin}, \texttt{compute}, and \texttt{copyout} block in the DSL is mapped to a corresponding AI Core function in AscendC, enforcing a clear execution structure and preventing invalid interleavings of computation and data movement.

We evaluate our approach on MultiKernelBench~\cite{wen2025multikernelbenchmultiplatformbenchmarkkernel} across seven operator categories.
Experimental results show that our method substantially improves both compilation success and functional correctness compared to direct LLM-based AscendC generation, achieving a \textbf{98.1\%} compilation success rate and an overall \textbf{90.4\%} functional correctness rate.
In addition, the generated kernels demonstrate competitive performance relative to PyTorch eager execution: \textbf{82.7\%} of kernels reach at least \textbf{20\%} of PyTorch eager performance, \textbf{57.7\%} reach \textbf{80\%}, and \textbf{46.2\%} match or exceed PyTorch eager baselines.
Beyond benchmark evaluation, we further validate \ourtool on two newly proposed real-world kernels from the mHC architecture~\cite{xie2026mhcmanifoldconstrainedhyperconnections}. \ourtool generates correct implementations for both kernels in a single pass, achieving \textbf{6.6$\times$} and \textbf{3.0$\times$} speedup over PyTorch eager execution, respectively. Starting from these generated kernels, a human developer collaboratively optimized them with the assistance of LLMs, iteratively refining performance. After optimization, the final implementations achieved up to \textbf{15.9$\times$} and \textbf{7.2$\times$} speedups over PyTorch eager execution, respectively. These results demonstrate that a carefully designed DSL, together with a structured and constrained transcompilation pipeline, not only enables effective LLM-based kernel generation for NPUs, but also provides a strong foundation for practical performance tuning on emerging workloads.

\section{Background and Motivation}
\subsection{Ascend NPU Architecture}
\begin{figure}[t]
  \centering
  \includegraphics[width=\linewidth]{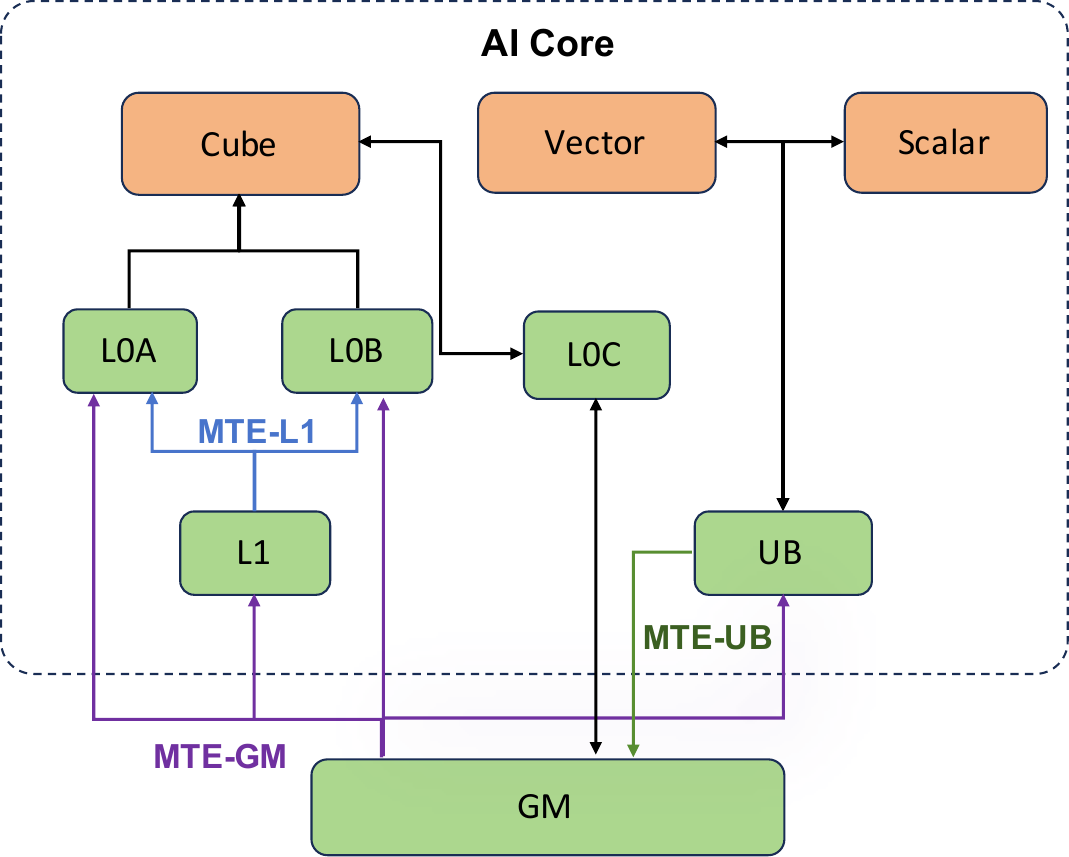}
  \caption{Ascend architecture.}
  \label{fig:ascend}
\end{figure}
Ascend~\cite{Ascend} is a domain-specific neural processing unit (NPU) designed from the ground up for DL workloads. Figure~\ref{fig:ascend} illustrates the overall architecture of the Ascend NPU.

\paragraph{Compute Units.}
The core computing component of Ascend is the \emph{AICore}, which integrates multiple heterogeneous compute units specialized for different types of operations. Specifically, each AICore consists of three types of units: Scalar Unit, Vector Unit, and Cube Unit.
The Scalar unit is responsible for scalar data processing and program control flow. The Vector unit executes vector operations, similar to traditional Single Instruction Multiple Data (SIMD) execution, where each vector instruction performs the same operation on multiple data elements simultaneously, such as element-wise arithmetic, normalization, activation functions, and reductions. The Cube unit is dedicated to matrix computations and can complete a matrix multiplication between matrix A (M × K) and matrix B (K × N) in a single execution. By allowing these units to operate in parallel, Ascend achieves high utilization across diverse operator patterns.

\paragraph{Memory Hierarchy.}
Ascend adopts a multi-level memory hierarchy consisting of Global Memory (GM), L1 buffers, unified buffers and L0 buffers. Ascend exposes its on-chip memory hierarchy to software, enabling explicit control over data placement and movement. This design provides greater flexibility for optimizing data locality and reuse~\cite{ascend_profile}.

\paragraph{Data Movement and Transfer Mechanism.}
Data transfers are managed by a dedicated hardware component called the Memory Transfer Engine (MTE). Ascend includes multiple MTE units, each responsible for transfers originating from a specific memory level. Transfers within the same MTE are executed sequentially, while transfers across different MTEs can proceed in parallel, enabling higher overall bandwidth utilization.

\paragraph{Instruction Pipelining.}
Ascend employs an explicit instruction pipelining mechanism to overlap computation and data movement. Both compute operations and memory transfers are issued as instructions and dispatched to separate execution queues corresponding to the Scalar, Vector, Cube, and MTE units. Instructions within a single queue are executed in order, while instructions in different queues can execute concurrently. This design enables fine-grained coordination between computation and data transfers, improving overall throughput when properly scheduled.

\subsection{AscendC Programming Model}

AscendC is a pipeline-based programming model designed for developing high-performance computational kernels on the Ascend NPU~\cite{wroblewski2025parallel}. Built on top of C++, AscendC exposes low-level hardware capabilities while providing structured abstractions that simplify kernel development.

\paragraph{Execution and Pipeline Model.}
AscendC follows a pipeline-based execution model that decomposes kernel execution into three logical stages: \emph{CopyIn}, \emph{Compute}, and \emph{CopyOut}. In the CopyIn stage, data is explicitly transferred from global memory to on-chip buffers via MTE operations. The Compute stage performs arithmetic operations using the Scalar, Vector, or Cube units. Finally, the CopyOut stage writes results back to global memory. This explicit separation enables instruction pipelining and overlap between data movement and computation, but requires careful coordination of buffer usage and data dependencies across stages.

Kernel execution is further organized into multiple \emph{blocks}, which represent the smallest logical execution units. The number of blocks is specified at kernel launch time, enabling parallel execution across multiple cores. AscendC also provides explicit synchronization primitives, such as \texttt{SyncAll}, to coordinate execution across blocks when necessary.

\paragraph{Tensor and Buffer Management.}
AscendC introduces tensor abstractions to represent data stored at different memory levels. \texttt{GlobalTensor} represents buffers allocated in global memory and serves as the interface for all kernel inputs and outputs. \texttt{LocalTensor} represents buffers allocated in on-chip memory, including Unified Buffer (UB), L1 buffers, and L0 buffers (e.g., L0A, L0B, and L0C).

While these abstractions closely reflect the underlying hardware, developers must explicitly manage buffer allocation and placement. In particular, on-chip buffers such as UB impose strict hardware constraints, including alignment requirements (e.g., 32-byte alignment) and size granularity restrictions. These constraints require careful reasoning about tensor shapes, tiling factors, and memory layouts, increasing the burden of kernel development.

\paragraph{Queues and Data Dependency Management.}
To coordinate data movement and computation across heterogeneous hardware units, AscendC provides a queue-based mechanism for managing data dependencies. After a hardware unit finishes operating on a tensor, the tensor pointer is enqueued. Subsequent stages dequeue the tensor, blocking if necessary until the data becomes available. This mechanism makes data dependencies explicit and avoids manual synchronization between pipeline stages.

Queues can hold multiple tensors, enabling techniques such as double buffering by increasing queue capacity. While powerful, effective use of queues requires careful design to ensure correct execution order and optimal overlap between data transfer and computation.

\paragraph{Operations and Programming Complexity.}
AscendC defines a rich set of operations corresponding to different hardware units, including data transfer operations such as \texttt{DataCopy}, matrix operations such as \texttt{Mmad} on the Cube unit, and vector operations such as \texttt{Adds} on the Vector unit. Constructing efficient kernels requires developers to carefully schedule these operations across pipeline stages, manage buffer reuse, and align instruction execution with hardware queues.

Overall, while AscendC provides expressive and powerful abstractions, writing correct and high-performance AscendC kernels demands deep hardware knowledge and meticulous coordination of memory allocation, data movement, and computation. These challenges motivate the need for higher-level abstractions and automated approaches for AscendC kernel generation, which we explore in this work.

\subsection{LLM for Kernel Generation}

LLMs have recently demonstrated strong capabilities in code generation and program synthesis~\cite{li2023starcodersourceyou,deepseekai2024deepseekcoderv2breakingbarrierclosedsource}, motivating their use in automating low-level system programming tasks~\cite{liu2025sharpen, VeriGen}. Prior work has explored LLM-based approaches for generating deep learning kernels, particularly for GPUs, where abundant public code, mature programming models, and well-documented optimization patterns provide rich training corpora.

\paragraph{LLMs for GPU Kernel Generation.}
Recent studies~\cite{baronio2025kevinmultiturnrlgenerating,li2025autotriton,woo2025tritonrl} have demonstrated that LLMs can effectively assist in generating GPU kernels in CUDA or Triton.
Prior work explores a range of paradigms, including direct code generation~\cite{ouyang2025kernelbenchllmswriteefficient}, multi-agent frameworks~\cite{lei2025pragma,du2025akg,zhu2025qimengkernelmacrothinkingmicrocodingparadigm}, and evolutionary optimization approaches~\cite{liao2025kernelevolve,guo2025evoengineer}.
These methods benefit from the rich ecosystem surrounding GPU programming, including abundant open-source kernels, extensive documentation, and standardized abstractions.
Consequently, LLMs are able to learn and reuse common optimization strategies—such as tiling, shared memory utilization, and instruction-level parallelism—from existing corpora.

\paragraph{Challenges in NPU Kernel Generation.}
In contrast, applying LLMs to NPU kernel generation presents distinct challenges. First, NPU programming models like AscendC impose specific constraints, such as strict memory alignment and complex synchronization across heterogeneous compute units. These demand explicit control over memory, data movement, and pipelines, raising code complexity and reducing error tolerance. Second, publicly available NPU kernel implementations are far scarcer than GPU counterparts, limiting the training corpus for LLMs. The lack of large-scale, high-quality NPU code makes it difficult for LLMs to generate correct and efficient kernels.

\paragraph{Limitations of Direct LLM-Based Code Generation.}
Due to these challenges, naively prompting LLMs to generate low-level NPU kernels often fails to produce even syntactically correct or compilable code~\cite{wen2025multikernelbenchmultiplatformbenchmarkkernel}. The structural complexity of kernels exacerbates LLM hallucinations and inconsistencies. Consequently, achieving reliable memory usage, synchronization, and compliance with hardware constraints—such as alignment and buffer placement—remains far from attainable in direct generation. These limitations highlight that a purely end-to-end, LLM-based approach is insufficient for generating functionally correct and performant NPU kernels.

\subsection{Motivation}
The limitations of direct LLM-based kernel generation motivate a shift toward intermediate representations that better align with the strengths of LLMs. While LLMs struggle to reliably emit low-level NPU code with strict structural and hardware constraints, they are considerably more effective at reasoning about higher-level algorithmic structure, dataflow, and tiling strategies. To bridge this gap between high-level intent and low-level NPU kernel code, recent work advocates introducing structured intermediate representations or domain-specific languages to guide LLM generation~\cite{liu2025sharpen,spec-kit}. By constraining the generation space and explicitly encoding hardware-relevant semantics at an appropriate level of abstraction, such representations improve both correctness and controllability. Following this direction, our approach leverages an LLM-guided DSL as an intermediate step for systematically generating AscendC kernels.

\section{DSL Definition}\label{sec:dsl_def}
\begin{figure}[t]
  \centering
  \includegraphics[width=\linewidth]{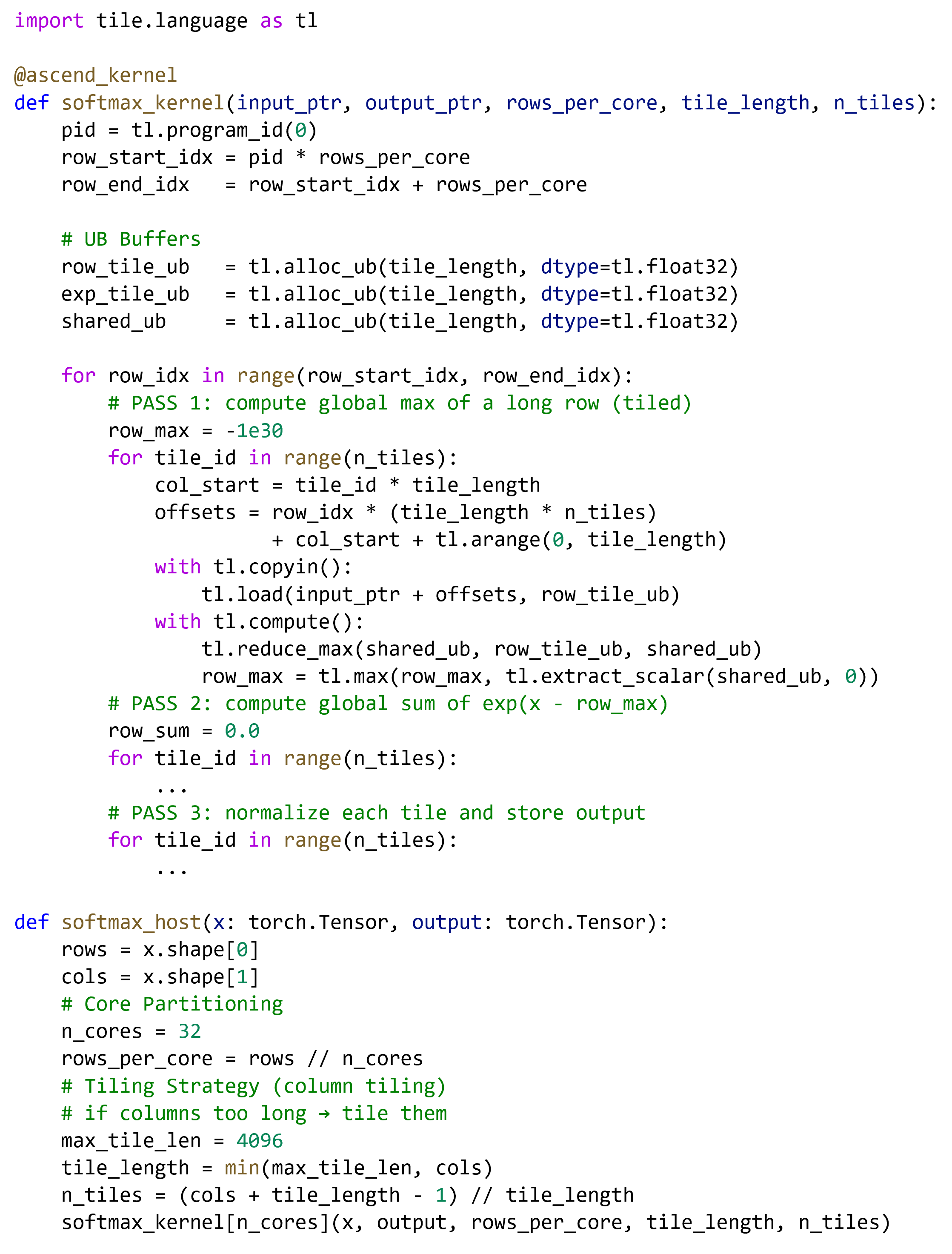}
  \caption{An example of a Softmax kernel written in the proposed Ascend DSL. The example shows the separation between host-side execution planning and kernel-side on-chip computation, as well as the staged execution pattern (\texttt{copyin}, \texttt{compute}, \texttt{copyout}).
It demonstrates how the DSL expresses tiled computation, on-chip buffer usage, and core dataflow logic while abstracting low-level AscendC details.}
  \label{fig:dsl}
\end{figure}
To bridge the gap between high-level kernel intent and low-level AscendC implementation, we introduce a lightweight domain-specific language (DSL) tailored for Ascend kernel generation.
The DSL is designed to be \emph{LLM-friendly}, \emph{appropriately abstracted}, and \emph{sufficiently expressive} to capture high-performance NPU kernel designs.
It provides a structured, hardware-aware representation that exposes core execution logic—such as tiling, dataflow, and on-chip execution structure—while shielding the LLM from unnecessary platform-specific complexity.

\paragraph{Overall Structure.}
An Ascend DSL program consists of two components: a \emph{host function} and a \emph{kernel function}, which respectively describe host-side execution control and on-chip computation.
This organization follows a widely used paradigm in accelerator programming and keeps each component concise and semantically focused.
Combined with a Triton-like programming style featuring compact syntax and clear structure, the DSL reduces unnecessary boilerplate and enables LLMs to reason about kernel execution at a higher level, focusing on core algorithmic decisions such as tiling, dataflow, and computation structure.

\paragraph{Host Function: Global Planning.}
The host function specifies global execution decisions, including core partitioning and tiling strategy.
Core partitioning defines how the overall workload is distributed across hardware cores by explicitly specifying the number of cores and the per-core workload assignment.
The DSL requires these decisions to be stated explicitly, making inter-core execution structure visible and analyzable.

The tiling strategy specifies how global tensors are partitioned into tiles that fit within on-chip memory (e.g., the Unified Buffer or L1).
All tiling parameters—such as tile length or block dimensions—must be explicitly defined, together with a brief rationale explaining the underlying memory constraints.
Finally, the host function launches the kernel and passes the computed tiling parameters to the kernel for on-chip execution.

\paragraph{Kernel Function: On-Chip Execution.}
The kernel function describes all on-chip behavior, including buffer allocation and computation.
All temporary on-chip buffers must be explicitly allocated within the kernel using dedicated DSL primitives for Unified Buffer (UB) or L1 memory.
The DSL disallows implicit aliasing and enforces explicit buffer declaration, ensuring that data movement and buffer reuse are transparent to both the LLM and the downstream transcompilation process.

Computation within the kernel follows a \emph{staged execution model}.
Global-to-on-chip data transfers must occur inside \texttt{copyin} blocks, computation must be enclosed within \texttt{compute} blocks, and on-chip-to-global transfers must be placed in \texttt{copyout} blocks.
Multiple \texttt{copyin}, \texttt{compute}, and \texttt{copyout} blocks are allowed, enabling the expression of multi-stage pipelines and iterative computation patterns.
This explicit staging captures Ascend-specific execution semantics while imposing a clear and regular structure that is easy for LLMs to generate and reason about.

The DSL provides a set of computation primitives covering common vector and reduction operations.
Their parameterization closely mirrors corresponding AscendC APIs, facilitating systematic and reliable translation during transcompilation.

\paragraph{Design Rationale.}
Overall, the DSL strikes a deliberate balance between abstraction and control.
It hides low-level AscendC details that impose significant programming burden—such as strict memory alignment handling and verbose \texttt{DataCopyPad} configurations—while explicitly modeling performance-critical aspects such as on-chip memory allocation, staged execution, and inter-core structure.
By doing so, the DSL reduces ambiguity in LLM-generated programs, enables structure-preserving transcompilation, and serves as an effective intermediate representation for generating correct and efficient AscendC kernels.

\section{Method}
\begin{figure*}[t]
  \centering
  \includegraphics[width=0.95\textwidth]{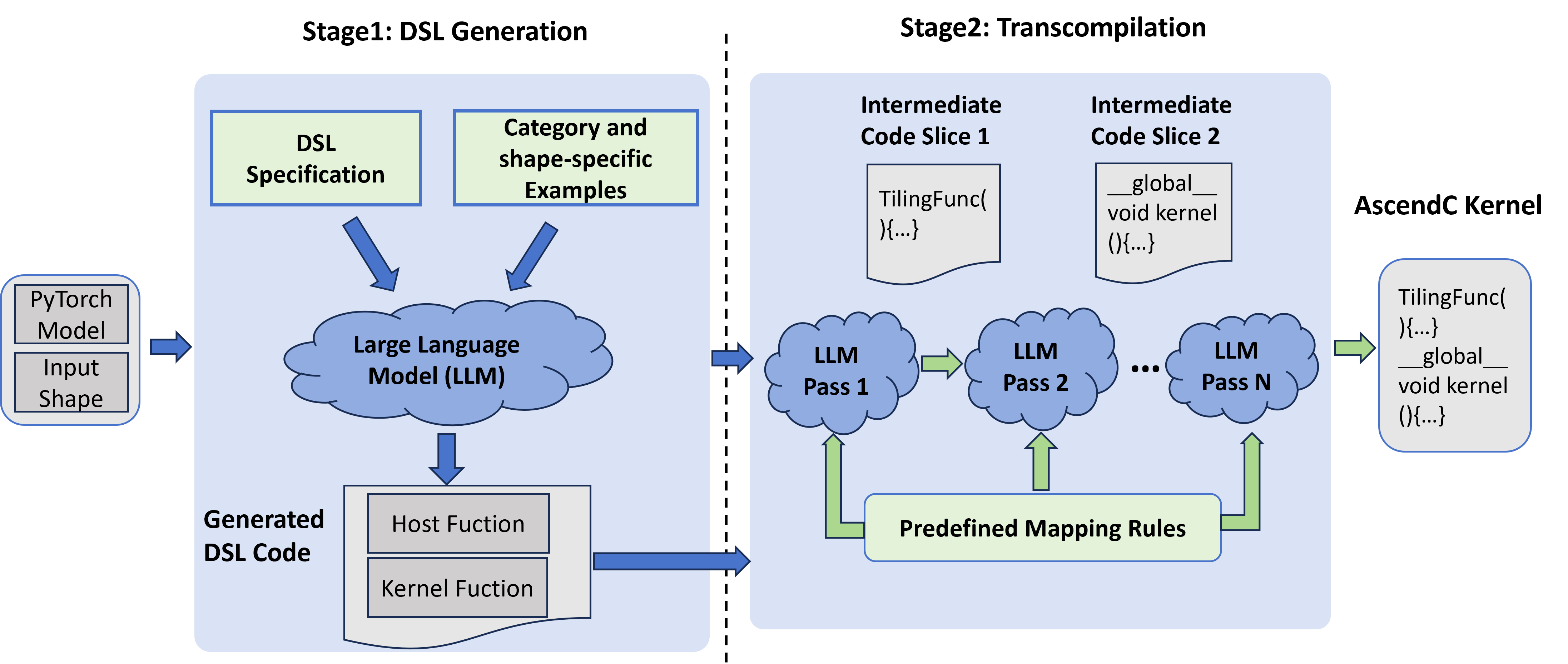}
  \caption{Framework of \ourtool.}
  \label{fig:overview}
\end{figure*}

Figure~\ref{fig:overview} illustrates our approach, \ourtool, for automatically generating AscendC kernels. \ourtool consists of two main stages: DSL generation and transcompilation. In the first stage, given a kernel task, the LLM generates DSL code that describes the kernel’s core logic, guided by the DSL specification and category-specific examples. In the second stage, the generated DSL is transcompiled into AscendC through multiple LLM passes, following predefined mapping rules in a step-by-step manner.
\subsection{DSL Code Generation}

The first stage of \ourtool is to generate DSL code that captures the core computation and execution strategy of an Ascend kernel. We leverage an LLM to produce DSL by carefully constructing prompts that constrain the generation space while preserving sufficient flexibility for optimization.

\paragraph{Prompt Design.}
The prompt for DSL code generation mainly consists of two components: (1) the DSL specification and (2) representative DSL examples. The DSL specification defines the syntax and semantics of the language, including the separation between host and kernel functions, explicit memory allocation, and the CopyIn--Compute--CopyOut execution structure. Due to the concise and structured design of the DSL, which is conceptually similar to Triton-style kernel descriptions, a compact specification is sufficient for the LLM to grasp the language rules and generate syntactically valid DSL programs.

\paragraph{Role of Category- and Shape-Specific Examples.}
In addition to the DSL specification, category- and shape-specific examples play a critical role in guiding DSL generation. These examples expose the LLM to optimization patterns that are closely tied to Ascend hardware characteristics, such as tiling strategies, buffer usage, and data movement patterns for different operator categories (e.g., element-wise, reduction, or matrix operations). By providing examples that match the target operator type and tensor shapes, the LLM can more effectively infer appropriate execution strategies and avoid implausible designs.

\paragraph{Effectiveness of the DSL Abstraction.}
Our DSL is intentionally designed to be concise yet expressive, capturing the essential patterns for high-performance Ascend kernels—such as explicit memory hierarchy usage and pipeline structure. This focused abstraction not only reduces the complexity of the code generation task, but also allows the LLM to more readily learn underlying computational and optimization principles from the provided examples. By abstracting away hardware-specific intricacies, the LLM can concentrate on understanding and expressing algorithmic intent and performance strategies. Empirically, this enables the model to generalize effectively from a limited set of examples and produce high-quality DSL code across diverse operator instances.

\subsection{Transcompilation}
\label{sec:transcompilation}

After obtaining the generated DSL program, the second stage of our approach transcompiles the DSL into executable AscendC code. Rather than directly generating complete AscendC kernels in a single step, we decompose the lowering process into a sequence of structured LLM-guided passes. This multi-pass design constrains generation at each stage, significantly improving correctness and robustness when targeting the highly restrictive AscendC programming model.

\paragraph{Overview of Multi-Pass Lowering.} The transcompilation process consists of four passes: (1) host-side translation, (2) kernel initialization translation, (3) kernel computation translation, and (4) an optional alignment and padding refinement pass. Each pass operates on a well-defined subset of the DSL semantics and produces a partially completed AscendC kernel, which is then consumed by the subsequent pass.

\paragraph{Prompt Design. }For each pass, the LLM is guided by a carefully constructed prompt that provides explicit context and constraints for the current translation stage. The prompt consists of four key components:
(1) explicit mapping rules used in this pass that specify how DSL elements are translated into AscendC code, for example, mapping DSL buffers to \texttt{TQue} or \texttt{TBuf} with designated execution roles and corresponding initialization patterns,
(2) relevant AscendC API information, including function signatures, buffer interfaces, and usage constraints,
(3) illustrative examples demonstrating the expected transformation, and
(4) the partially generated AscendC code produced by the previous pass.

\paragraph{Pass 1: Host-Side Translation.}
The first pass translates the host-side portion of the DSL into AscendC host code. This includes defining tiling-related data structures, computing tiling parameters (e.g., per-core workload size and tile dimensions), and configuring these parameters using AscendC host APIs. The host code is responsible for computing all execution parameters required by the kernel and launching the kernel with the appropriate number of blocks.

\paragraph{Pass 2: Kernel Initialization.}
The second pass generates the kernel initialization logic, which prepares all state required before entering the computation loop. Tiling parameters computed on the host are copied into kernel member variables. Each kernel instance retrieves its block identifier using \texttt{GetBlockIdx()}, which is used to compute the global memory offsets corresponding to the core’s assigned data partition.

This pass also initializes on-chip memory resources. Based on the DSL buffer annotations, buffers used for data transfer are mapped to AscendC tensor queues (\texttt{TQue}), while temporary working buffers are mapped to tensor buffers (\texttt{TBuf}). Input and output queues are initialized with appropriate capacities to support pipelining and, when needed, double buffering.

\paragraph{Pass 3: Kernel Computation Translation.}
The third pass performs the core translation of DSL computation logic into AscendC kernel code. As defined in Section~\ref{sec:dsl_def}, the DSL enforces a strict separation between \emph{CopyIn}, \emph{Compute}, and \emph{CopyOut} stages. We preserve this structure by translating each DSL block into a corresponding AscendC function.

Specifically, the kernel’s \texttt{Process()} method implements the per-core execution loop and invokes a sequence of stage functions, such as \texttt{CopyInX}, \texttt{ComputeX}, and \texttt{CopyOutX}. Each stage function is defined as an \texttt{\_\_aicore\_\_ inline} function, optionally parameterized by the loop index for global memory address calculation.

Within \texttt{CopyIn} functions, AscendC \texttt{DataCopy} operations are used to transfer data from global memory to on-chip buffers, after which the data is enqueued to the appropriate input queues. In \texttt{Compute} functions, all required input tensors are dequeued at the beginning of the function, temporary buffers are accessed through pre-initialized \texttt{TBuf} objects, and DSL compute operations are mapped to AscendC APIs. Results are enqueued to output queues or forwarded to subsequent compute stages as needed. Finally, \texttt{CopyOut} functions dequeue result tensors and write them back to global memory using \texttt{DataCopy}. Synchronization primitives such as \texttt{SyncAll} are inserted only when cross-kernel data dependencies are required.

By translating each DSL block into a separate function with a fixed role, this pass strictly constrains the structure of the generated kernel and prevents illegal interleavings of data movement and computation.

\paragraph{Pass 4: Alignment and Padding Refinement.}
The final pass is optional and addresses hardware-specific edge cases related to memory alignment and non-uniform tensor shapes. AscendC imposes strict alignment requirements on data transfers (e.g., 32-byte alignment for UB accesses). When DSL-level tiling or tensor shapes do not naturally satisfy these constraints, this pass replaces standard \texttt{DataCopy} operations with \texttt{DataCopyPad} and configures additional parameters such as padding size, stride, and layout transformation. This refinement pass improves robustness without complicating earlier lowering stages.

\paragraph{Per-Pass Correction Feedback.}
Our multi-pass transcompilation process ensures that each lowering pass produces compilable AscendC code.
After every pass, the generated code is compiled to obtain concrete compiler feedback.
If compilation errors occur, the error messages are fed back to the LLM, which is prompted to revise and fix the code before proceeding to the next pass.

\paragraph{Benefits of Staged Transcompilation.}
Overall, decomposing transcompilation into multiple LLM-guided passes allows each step to focus on a narrow, well-defined task. This design reduces hallucination, improves structural correctness, and ensures that the generated AscendC kernels adhere to both the semantic constraints of the DSL and the low-level requirements of the AscendC programming model.

\section{Evaluation}

We seek to answer the following research questions:

\noindent \textbf{RQ1. }Can \ourtool generate functionally correct AscendC kernels?

\noindent \textbf{RQ2. }How does the performance of kernels generated by \ourtool compare to Pytorch eager execution baseline?

\noindent \textbf{RQ3. }Can \ourtool generalize to newly proposed operators outside existing benchmarks and support practical kernel development workflows?
\subsection{Experimental Setup}

\paragraph{Benchmark.}
We evaluate our approach using \textit{MultiKernelBench}~\cite{wen2025multikernelbenchmultiplatformbenchmarkkernel}, a multi-platform kernel benchmark suite that supports AscendC, CUDA, and other accelerator backends. MultiKernelBench extends \textit{KernelBench}~\cite{ouyang2025kernelbenchllmswriteefficient} by unifying kernel interfaces across heterogeneous platforms. Following prior work, we adopt the Level 1 task categories from KernelBench, which consist of single-operator kernels with moderate complexity and well-defined computation and memory access patterns. These tasks cover a diverse set of operator types, including activation, reduction, normalization, and mathematical operations, and are suitable for evaluating both functional correctness and performance of automatically generated kernels. 

Notably, we use the updated task shapes from the latest KernelBench release~\cite{kernelbenchNewShapeBlog}, which scale tensor sizes to ensure kernel execution times exceed 15\,ms.
This setting leads to a more realistic and challenging evaluation: kernel launch overhead no longer dominates the measured runtime, making performance comparisons more scientifically meaningful.
As a result, achieving \textbf{Fast$_{0.8}$} or \textbf{Fast$_{0.2}$} under this configuration is substantially more difficult than in earlier benchmark settings.

\paragraph{Environment.}
All experiments are conducted on an Ascend 910B2 NPU. We use the CANN toolkit version 8.1 and PyTorch 2.6 as the host framework. The experimental platform runs Ubuntu 22.04 with official Ascend drivers and firmware versions matched to the CANN release. All kernels are compiled using the default Ascend compiler toolchain.

\paragraph{Reported Metrics.}
We report the following metrics to evaluate both the correctness and performance of generated AscendC kernels:
(i) \textbf{Compilation Rate}, the percentage of generated kernels that successfully compile without errors;
(ii) \textbf{Correctness Rate (Pass@1)}, the percentage of compiled kernels that produce numerically correct outputs compared with reference implementations;
and (iii) \textbf{Performance Metrics}, including \textbf{Fast$_{0.2}$}, \textbf{Fast$_{0.8}$}, and \textbf{Fast$_{1.0}$}.
Fast$_x$ measures the percentage of correct kernels whose execution time is within $x \times$ of the performance of
reference implementations.

Fast$_{0.2}$ captures whether the generated kernel satisfies basic performance requirements, such as leveraging
vectorized execution instead of scalar operations and utilizing multiple AI cores.
Fast$_{0.8}$ indicates that the kernel adopts a reasonably efficient algorithm and dataflow strategy suitable for
practical deployment.
Fast$_{1.0}$ denotes kernels whose performance is on par with or better than the reference implementations.

\subsection{RQ1: Correctness}
\begin{table}[t]
\caption{Correctness evaluation of operator implementations by category.}
\label{table:correctness}
\centering
\begin{tabular}{lll}
\toprule
\textbf{Kernel Category}  & \textbf{Comp@1} & \textbf{Pass@1}    \\
\midrule
Activation (15 kernels)         &    100.0      &  100.0\\
Loss (7 kernels)              &    100.0       &  85.7\\
Math (6 kernels)             &    83.3                  & 83.3 \\
Normalization (8 kernels)      &    100.0      & 87.5 \\
Optimizer  (5 kernels)        &    100.0  & 100.0 \\
Reduce (5 kernels)            &    100.0       &  100.0\\
Pooling (6 kernels) &100.0 & 66.7\\

\midrule
\textbf{Total (52 kernels)}&98.1&90.4 \\
\bottomrule
\end{tabular}
\footnotemark
\end{table}

\footnotetext{MatMul and Convolution kernels are not included in this evaluation and are part of ongoing work. AscendC Cube programming provides both high-level and low-level interfaces. The low-level interface is highly complex, while the high-level interface differs substantially from the standard AscendC programming model, lacking explicit \texttt{copyin}, \texttt{compute}, and \texttt{copyout} stages as well as queue-based execution. To address these challenges, we are exploring the use of CATLASS~\cite{CATLASS} as an alternative target language for these kernels.}

Table~\ref{table:correctness} summarizes the results across different operator categories from MultiKernelBench. Overall, \ourtool achieves a compilation success rate of \textbf{98.1\%} and a functional correctness rate of \textbf{90.4\%} across a total of 52 kernels. This represents a substantial improvement over prior direct LLM-based AscendC generation approaches~\cite{wen2025multikernelbenchmultiplatformbenchmarkkernel}, which exhibit very low end-to-end correctness—for example, Claude Sonnet-4 achieves only \textbf{13.0\%} correctness on the same benchmark~\cite{wen2025multikernelbenchmultiplatformbenchmarkkernel}.

Across operator categories, most kernels achieve near-perfect compilation rates. Activation, Normalization, Optimizer, Reduce, and Pooling operators all reach \textbf{100\% Comp@1}, demonstrating that \ourtool can reliably generate syntactically valid AscendC code for a wide range of common operator patterns. The slightly lower compilation rate observed for Math operators is primarily due to the \texttt{mask\_cumsum} kernel, which involves boolean data types that are not yet fully covered by the current prompt knowledge.

In terms of functional correctness, performance varies across categories. Activation, Optimizer, and Reduce operators achieve \textbf{100\% Pass@1}, benefiting from well-structured dataflow patterns and clear reduction semantics. Normalization operators also show strong correctness, reaching \textbf{87.5\% Pass@1}. Pooling operators exhibit relatively lower Pass@1, which we attribute to their more complex control flow and boundary-sensitive computations, making them more prone to subtle logic errors during generation.

Overall, these results demonstrate that the combination of a structured DSL, category-specific exemplars, and multi-pass transcompilation with explicit structural constraints can significantly improve the correctness of automatically generated AscendC kernels, even across a diverse set of operator types.

\subsection{RQ2: Performance}
\begin{table}[t]
\caption{Performance evaluation of generated AscendC kernels by operator category.}
\label{table:performance}
\centering
\begin{tabular}{lccc}
\toprule
\textbf{Kernel Category} 
& \textbf{Fast$_{0.2}$@1} 
& \textbf{Fast$_{0.8}$@1} 
& \textbf{Fast$_{1.0}$@1} \\
\midrule
Activation        &  100.0 &  80.0 &  40.0 \\
Loss              &  85.7 &  85.7 &  85.7 \\
Math              & 83.3  & 66.7  &  66.7 \\
Normalization     &  50.0 &  37.5 &  37.5 \\
Optimizer         &  100.0 &  100.0 &  100.0 \\
Reduce            &  100.0 &  0.0 &  0.0 \\
Pooling &50.0&0.0&0.0\\
\midrule
\textbf{Total}    &  82.7 &  57.7 &  46.2 \\
\bottomrule
\end{tabular}
\end{table}

Table~\ref{table:performance} summarizes the performance of the generated AscendC kernels.
\textbf{Fast$_{\alpha}$} denotes the percentage of kernels whose execution time reaches at least $\alpha \times$ that of the PyTorch eager execution baseline, with $\alpha \in {0.2, 0.8, 1.0}$.

Overall, \ourtool produces a large fraction of kernels with competitive performance.
Across all operator categories, \textbf{82.7\%} of kernels achieve at least \textbf{20\%} of PyTorch eager performance, \textbf{57.7\%} reach \textbf{80\%}, and \textbf{46.2\%} match or exceed the eager baseline.
These results demonstrate that the proposed DSL-guided transcompilation approach is able to recover effective dataflow organization and tiling strategies for a wide range of operators, even without manual tuning.

Performance varies across operator categories. Activation, Optimizer and Loss operators exhibit strong performance, where much of the observed speedup is attributed to effective operator fusion. Math operators also achieve high performance under relaxed thresholds, with a substantial portion of kernels even surpassing PyTorch eager execution. Notably, for several Math kernels, the performance gains are not due to fusion but instead stem from genuine kernel optimizations tailored to the computation patterns of these operators.

Normalization operators exhibit lower performance, particularly under stricter thresholds. These operators often involve multi-stage computations and intermediate reductions, making their performance more sensitive to memory layout, execution ordering, and reduction strategies. While \ourtool is able to generate correct and functional implementations for these operators, achieving peak performance may require more fine-grained, expert-written examples that expose hardware-specific optimization patterns.

For Reduce operators, we observe that performance improvements largely depend on low-level instruction control and reduction primitives, which are difficult to express purely at the DSL level. This suggests that certain optimizations remain challenging to infer from high-level abstractions alone. These observations highlight promising directions for future work, including automatic optimization across both the DSL level and the generated AscendC code. Importantly, \ourtool already provides a strong and correct starting point, upon which further performance tuning can be systematically applied. 

Overall, the results demonstrate that \ourtool can automatically generate performant AscendC kernels for many common operator categories, while also revealing opportunities for further optimization on more complex and performance-sensitive workloads.

\subsection{RQ3: Practical Usage}

To evaluate the practical applicability of \ourtool, we apply it to a real and recently proposed kernel development task.
Specifically, we consider the Manifold-Constrained Hyper-Connections (mHC) introduced in DeepSeek~\cite{xie2026mhcmanifoldconstrainedhyperconnections}, which represents a newly proposed model structure not covered by existing benchmark suites.

We focus on two kernels, \texttt{mHC\_post} and \texttt{mHC\_post\_grad}.
To evaluate our approach, we first implement their reference behavior using PyTorch APIs and specify representative input shapes as task definitions.
These specifications are then provided to \ourtool, which automatically generates the corresponding DSL programs and transcompiles them into AscendC code.

Despite the novelty of this kernel, \ourtool is able to generate functionally correct AscendC implementations in a single pass.
Under the evaluated input shapes, the generated kernels achieve a \textbf{6.6$\times$} speedup over PyTorch eager execution for \texttt{mHC\_post}, and a \textbf{3.0$\times$} speedup for \texttt{mHC\_post\_grad}.
These results demonstrate that the proposed DSL and transcompilation pipeline generalize beyond benchmark operators and can effectively handle emerging workloads with complex computation patterns.

Starting from the generated kernels, we further invited an experienced AscendC developer to perform optimization with the assistance of LLMs.
Thanks to the structured and readable code produced by \ourtool, the developer was able to complete optimization for both kernels within one day, with optimization strategies expressed in natural language and translated into code by the LLM.
Without such a structured starting point, directly optimizing raw AscendC code using LLMs would be extremely challenging, as current models struggle to generate correct low-level kernel implementations from scratch~\cite{wen2025multikernelbenchmultiplatformbenchmarkkernel}.
After optimization, the final implementations achieved up to \textbf{15.9$\times$} and \textbf{7.2$\times$} speedup over PyTorch eager execution, respectively. 

This case study highlights two important aspects of \ourtool.
First, it can serve as a practical tool for rapidly bootstrapping correct NPU kernels for newly emerging operators.
Second, the generated kernels provide a high-quality starting point for further expert-level optimization, significantly reducing development cost while retaining strong performance potential.

\section{Conclusion}

This paper presents \ourtool, a DSL-guided framework for automatic AscendC kernel generation using large language models.
By introducing a concise and NPU-aware DSL and a structured, multi-pass transcompilation pipeline, \ourtool effectively bridges the gap between high-level kernel intent and low-level NPU execution semantics.
The proposed design enables LLMs to focus on core algorithmic decisions while avoiding the complexity and brittleness of directly generating platform-specific kernel code.

Experimental results on MultiKernelBench demonstrate that \ourtool substantially improves both compilation success and functional correctness compared to direct LLM-based AscendC generation, while producing performance-competitive kernels relative to PyTorch eager execution. Across a diverse set of operator categories, a significant fraction of the generated kernels match or exceed eager baselines, highlighting the effectiveness of the proposed DSL abstractions and structured transcompilation. Moreover, experiments on newly proposed mHC architecture show that \ourtool can generalize beyond benchmark workloads, generating correct kernels that significantly surpass PyTorch eager performance, further validating its practicality for real-world NPU kernel development.

Overall, \ourtool shows that combining appropriate abstractions with structured generation can make LLM-based kernel synthesis practical for NPUs.
Future work will focus on improving performance on more complex and performance sensitive operators, as well as exploring joint optimization across both the DSL and generated AscendC code to further close the gap with expert-tuned implementations.

\bibliographystyle{ACM-Reference-Format}
\bibliography{reference}

@INPROCEEDINGS{Ascend,
  author={Liao, Heng and Tu, Jiajin and Xia, Jing and Liu, Hu and Zhou, Xiping and Yuan, Honghui and Hu, Yuxing},
  booktitle={2021 IEEE International Symposium on High-Performance Computer Architecture (HPCA)}, 
  title={Ascend: a Scalable and Unified Architecture for Ubiquitous Deep Neural Network Computing : Industry Track Paper}, 
  year={2021},
  volume={},
  number={},
  pages={789-801},
  doi={10.1109/HPCA51647.2021.00071}}

@inproceedings{
wroblewski2025parallel,
title={Parallel Scan on Ascend {AI} Accelerators},
author={Bart{\l}omiej Wr{\'o}blewski and Gioele Gottardo and Anastasios Zouzias},
booktitle={Greeks in AI Symposium 2025},
year={2025},
url={https://openreview.net/forum?id=wPepcNWMhs}
}

@inproceedings{ascend_profile,
author = {Zhou, Yuhang and Wang, Zhibin and Liu, Guyue and Li, Shipeng and Lin, Xi and Wang, Zibo and Wang, Yongzhong and Wei, Fuchun and Zhang, Jingyi and Hu, Zhiheng and Liu, Yanlin and Li, Chunsheng and Zhang, Ziyang and Wang, Yaoyuan and Zhou, Bin and Dou, Wanchun and Chen, Guihai and Tian, Chen},
title = {Squeezing Operator Performance Potential for the Ascend Architecture},
year = {2025},
isbn = {9798400710797},
publisher = {Association for Computing Machinery},
address = {New York, NY, USA},
url = {https://doi.org/10.1145/3676641.3716243},
doi = {10.1145/3676641.3716243},
booktitle = {Proceedings of the 30th ACM International Conference on Architectural Support for Programming Languages and Operating Systems, Volume 2},
pages = {1156–1171},
numpages = {16},
location = {Rotterdam, Netherlands},
series = {ASPLOS '25}
}

@misc{wen2025multikernelbenchmultiplatformbenchmarkkernel,
      title={MultiKernelBench: A Multi-Platform Benchmark for Kernel Generation}, 
      author={Zhongzhen Wen and Yinghui Zhang and Zhong Li and Zhongxin Liu and Linna Xie and Tian Zhang},
      year={2025},
      eprint={2507.17773},
      archivePrefix={arXiv},
      primaryClass={cs.DC},
      url={https://arxiv.org/abs/2507.17773}, 
}

@article{liu2025sharpen,
  title={Sharpen the Spec, Cut the Code: A Case for Generative File System with SYSSPEC},
  author={Liu, Qingyuan and Mo, Zou and Zhang, Hengbin and Du, Dong and Xia, Yubin and Chen, Haibo},
  journal={arXiv preprint arXiv:2512.13047},
  year={2025}
}

@misc{ouyang2025kernelbenchllmswriteefficient,
      title={KernelBench: Can LLMs Write Efficient GPU Kernels?}, 
      author={Anne Ouyang and Simon Guo and Simran Arora and Alex L. Zhang and William Hu and Christopher Ré and Azalia Mirhoseini},
      year={2025},
      eprint={2502.10517},
      archivePrefix={arXiv},
      primaryClass={cs.LG},
      url={https://arxiv.org/abs/2502.10517}, 
}

@misc{baronio2025kevinmultiturnrlgenerating,
      title={Kevin: Multi-Turn RL for Generating CUDA Kernels}, 
      author={Carlo Baronio and Pietro Marsella and Ben Pan and Simon Guo and Silas Alberti},
      year={2025},
      eprint={2507.11948},
      archivePrefix={arXiv},
      primaryClass={cs.LG},
      url={https://arxiv.org/abs/2507.11948}, 
}

@article{wei2025astra,
  title={Astra: A multi-agent system for gpu kernel performance optimization},
  author={Wei, Anjiang and Sun, Tianran and Seenichamy, Yogesh and Song, Hang and Ouyang, Anne and Mirhoseini, Azalia and Wang, Ke and Aiken, Alex},
  journal={arXiv preprint arXiv:2509.07506},
  year={2025}
}

@online{kernelbenchNewShapeBlog,
  title={KernelBench v0.1},
  url={https://scalingintelligence.stanford.edu/blogs/kernelbenchv01/},
  year={2025}
}

@online{CATLASS,
  title={CATLASS},
  url={https://gitcode.com/cann/catlass},
  year={2026}
}

@online{spec-kit,
  title={Spec Kit},
  url={https://github.github.io/spec-kit/},
  year={2026}
}

@misc{li2023starcodersourceyou,
      title={StarCoder: may the source be with you!}, 
      author={Raymond Li and Loubna Ben Allal and Yangtian Zi and Niklas Muennighoff and Denis Kocetkov and Chenghao Mou and Marc Marone and Christopher Akiki and Jia Li and Jenny Chim and Qian Liu and Evgenii Zheltonozhskii and Terry Yue Zhuo and Thomas Wang and Olivier Dehaene and Mishig Davaadorj and Joel Lamy-Poirier and João Monteiro and Oleh Shliazhko and Nicolas Gontier and Nicholas Meade and Armel Zebaze and Ming-Ho Yee and Logesh Kumar Umapathi and Jian Zhu and Benjamin Lipkin and Muhtasham Oblokulov and Zhiruo Wang and Rudra Murthy and Jason Stillerman and Siva Sankalp Patel and Dmitry Abulkhanov and Marco Zocca and Manan Dey and Zhihan Zhang and Nour Fahmy and Urvashi Bhattacharyya and Wenhao Yu and Swayam Singh and Sasha Luccioni and Paulo Villegas and Maxim Kunakov and Fedor Zhdanov and Manuel Romero and Tony Lee and Nadav Timor and Jennifer Ding and Claire Schlesinger and Hailey Schoelkopf and Jan Ebert and Tri Dao and Mayank Mishra and Alex Gu and Jennifer Robinson and Carolyn Jane Anderson and Brendan Dolan-Gavitt and Danish Contractor and Siva Reddy and Daniel Fried and Dzmitry Bahdanau and Yacine Jernite and Carlos Muñoz Ferrandis and Sean Hughes and Thomas Wolf and Arjun Guha and Leandro von Werra and Harm de Vries},
      year={2023},
      eprint={2305.06161},
      archivePrefix={arXiv},
      primaryClass={cs.CL},
      url={https://arxiv.org/abs/2305.06161}, 
}

@misc{deepseekai2024deepseekcoderv2breakingbarrierclosedsource,
      title={DeepSeek-Coder-V2: Breaking the Barrier of Closed-Source Models in Code Intelligence}, 
      author={DeepSeek-AI and Qihao Zhu and Daya Guo and Zhihong Shao and Dejian Yang and Peiyi Wang and Runxin Xu and Y. Wu and Yukun Li and Huazuo Gao and Shirong Ma and Wangding Zeng and Xiao Bi and Zihui Gu and Hanwei Xu and Damai Dai and Kai Dong and Liyue Zhang and Yishi Piao and Zhibin Gou and Zhenda Xie and Zhewen Hao and Bingxuan Wang and Junxiao Song and Deli Chen and Xin Xie and Kang Guan and Yuxiang You and Aixin Liu and Qiushi Du and Wenjun Gao and Xuan Lu and Qinyu Chen and Yaohui Wang and Chengqi Deng and Jiashi Li and Chenggang Zhao and Chong Ruan and Fuli Luo and Wenfeng Liang},
      year={2024},
      eprint={2406.11931},
      archivePrefix={arXiv},
      primaryClass={cs.SE},
      url={https://arxiv.org/abs/2406.11931}, 
}

@article{VeriGen,
author = {Thakur, Shailja and Ahmad, Baleegh and Pearce, Hammond and Tan, Benjamin and Dolan-Gavitt, Brendan and Karri, Ramesh and Garg, Siddharth},
title = {VeriGen: A Large Language Model for Verilog Code Generation},
year = {2024},
issue_date = {May 2024},
publisher = {Association for Computing Machinery},
address = {New York, NY, USA},
volume = {29},
number = {3},
issn = {1084-4309},
url = {https://doi.org/10.1145/3643681},
doi = {10.1145/3643681},
journal = {ACM Trans. Des. Autom. Electron. Syst.},
month = apr,
articleno = {46},
numpages = {31}
}

@article{lei2025pragma,
  title={PRAGMA: A Profiling-Reasoned Multi-Agent Framework for Automatic Kernel Optimization},
  author={Lei, Kelun and Yang, Hailong and Zhang, Huaitao and You, Xin and Zhang, Kaige and Luan, Zhongzhi and Liu, Yi and Qian, Depei},
  journal={arXiv preprint arXiv:2511.06345},
  year={2025}
}

@article{du2025akg,
  title={AKG kernel Agent: A Multi-Agent Framework for Cross-Platform Kernel Synthesis},
  author={Du, Jinye and Yuan, Quan and Zhang, Zuyao and Yi, Yanzhi and Hu, Jiahui and Chen, Wangyi and Zhu, Yiyang and Zheng, Qishui and Zou, Wenxiang and Chang, Xiangyu and others},
  journal={arXiv preprint arXiv:2512.23424},
  year={2025}
}

@article{li2025autotriton,
  title={Autotriton: Automatic triton programming with reinforcement learning in llms},
  author={Li, Shangzhan and Wang, Zefan and He, Ye and Li, Yuxuan and Shi, Qi and Li, Jianling and Hu, Yonggang and Che, Wanxiang and Han, Xu and Liu, Zhiyuan and others},
  journal={arXiv preprint arXiv:2507.05687},
  year={2025}
}

@article{woo2025tritonrl,
  title={Tritonrl: Training llms to think and code triton without cheating},
  author={Woo, Jiin and Zhu, Shaowei and Nie, Allen and Jia, Zhen and Wang, Yida and Park, Youngsuk},
  journal={arXiv preprint arXiv:2510.17891},
  year={2025}
}

@article{liao2025kernelevolve,
  title={KernelEvolve: Scaling Agentic Kernel Coding for Heterogeneous AI Accelerators at Meta},
  author={Liao, Gang and Qin, Hongsen and Wang, Ying and Golden, Alicia and Kuchnik, Michael and Yetim, Yavuz and Ang, Jia Jiunn and Fu, Chunli and He, Yihan and Hsia, Samuel and others},
  journal={arXiv preprint arXiv:2512.23236},
  year={2025}
}

@article{guo2025evoengineer,
  title={EvoEngineer: Mastering Automated CUDA Kernel Code Evolution with Large Language Models},
  author={Guo, Ping and Zhu, Chenyu and Chen, Siyuan and Liu, Fei and Lin, Xi and Lu, Zhichao and Zhang, Qingfu},
  journal={arXiv preprint arXiv:2510.03760},
  year={2025}
}

@misc{zhu2025qimengkernelmacrothinkingmicrocodingparadigm,
      title={QiMeng-Kernel: Macro-Thinking Micro-Coding Paradigm for LLM-Based High-Performance GPU Kernel Generation}, 
      author={Xinguo Zhu and Shaohui Peng and Jiaming Guo and Yunji Chen and Qi Guo and Yuanbo Wen and Hang Qin and Ruizhi Chen and Qirui Zhou and Ke Gao and Yanjun Wu and Chen Zhao and Ling Li},
      year={2025},
      eprint={2511.20100},
      archivePrefix={arXiv},
      primaryClass={cs.DC},
      url={https://arxiv.org/abs/2511.20100}, 
}

@misc{xie2026mhcmanifoldconstrainedhyperconnections,
      title={mHC: Manifold-Constrained Hyper-Connections}, 
      author={Zhenda Xie and Yixuan Wei and Huanqi Cao and Chenggang Zhao and Chengqi Deng and Jiashi Li and Damai Dai and Huazuo Gao and Jiang Chang and Kuai Yu and Liang Zhao and Shangyan Zhou and Zhean Xu and Zhengyan Zhang and Wangding Zeng and Shengding Hu and Yuqing Wang and Jingyang Yuan and Lean Wang and Wenfeng Liang},
      year={2026},
      eprint={2512.24880},
      archivePrefix={arXiv},
      primaryClass={cs.CL},
      url={https://arxiv.org/abs/2512.24880}, 
}

\end{document}